# SOLARIS SYSTEM RESOURCE MANAGER: ALL I EVER WANTED WAS MY UNFAIR ADVANTAGE (AND WHY YOU CAN'T GET IT!)

Neil J. Gunther

*Performance Dynamics Consulting™*
*Castro Valley, California USA*
www.perfdynamics.com

Traditional UNIX time-share schedulers attempt to be fair to all users by employing a round-robin style algorithm for allocating CPU time. Unfortunately, a loophole exists whereby the scheduler can be biased in favor of a greedy user running many short CPU-time processes. This loophole is not a defect but an intrinsic property of the round-robin scheduler that ensures responsiveness to the short CPU demands associated with multiple interactive users. A new generation of UNIX system resource management (SRM) software constrains the scheduler to be equitable to all users regardless of the number of processes each may be running. This "fair-share" scheduling draws on the concept of pro rating resource "shares" across users and groups and then dynamically adjusting CPU usage to meet those share proportions. The simple notion of statically allocating these shares, however, belies the potential consequences for performance as measured by user response time and service level targets. We demonstrate this point by modeling several simple share allocation scenarios and analyzing the corresponding performance effects. A brief comparison of commercial SRM implementations from HEWLETT-PACKARD, IBM, and SUN MICROSYSTEMS is also presented.

## 1. INTRODUCTION

When President Lyndon B. Johnson was asked what made him such a skillful negotiator, he replied, "All I ever want is my unfair advantage!" [Beschloss 98]. Similarly for shared computational resources. If users *can* usurp resources to get their work done, they *will*. And not always by negotiating with other users, either. For enterprise level resources, such as scalable servers and networks, the opportunity for greedy resource consumption can present severe problems for system administration and capacity planning. It is quite impractical for a human administrator to hover over servers and networks watching for potentially selfish resource consumption. The only way around this kind of performance threat is adaptive control of system resources. Adaptive control algorithms for operating systems and data networks have been the subject of extensive discussions in the academic literature (see e.g., [Larmouth 75], [Henry 84], [Kay and Lauder 1988], [Greenberg 90]).

Only in recent years have reliable implementations of such algorithms begun to appear in commercially engineered systems e.g., IBM SRM [Samson 97] Workload Manager [IBM 94], some flavors of UNIX [Potter et al. 92], [Bettison 91] and network management systems [Keshav 97]. In this paper, we discuss capacity planning for one of the newest implementations called Solaris® Systems Resource Manager (SRM) from the standpoint of performance and capacity planning.

Although the *set it and forget it* approach of SRM is desirable for server capacity administration, dynamically changing SRM share allocations can have quite unintuitive consequences for user performance. A system administrator needs to understand the performance implications in order to make a judicious choice of SRM allocations. But heretofore the only way to gain such an understanding was by playing arbitrarily with SRM tunables. As far as we are aware, this is the first paper to present Solaris SRM principles of operation in a transparent way, along with allocation guidelines for performance based on an SRM capacity model.

This paper introduces the reader to SRM capacity planning and share allocation techniques for groups and users. A comparison with IBM's WorkLoad Manager (WLM) is presented at the end. Companion webpages by [McDougall 1999a and 1999b] present the underlying SRM modifications to the traditional UNIX scheduler [Bach 1986], [Ritchie & Thompson 1978], [Vahalia 1998], and the corresponding SRM tuning parameters. Although Solaris SRM can be set up to provide constraints on the usage of various server resources (e.g., CPU, memory, I/O) the focus in this paper is entirely on CPU scheduling since it is the most sensitive and dynamic of server resources. We commence by defining some important terminology.

## 2. WHAT IS 'FAIR?'

In the context of scheduling computer resources and routing packets, the term *fair* is meant to imply *equity* (not *equality)* in resource consumption. *SRM is equitable but not egalitarian*. The usual analogy lies with equity in a corporation. Executive management holds more equity in a company than underlings. Accordingly, the larger shareholders are entitled to a greater percentage of company profits. Similarly, under SRM, the more equity you hold (via allocated shares), the greater the percentage of server resources you are guaranteed at runtime. In SRM there are, however, some significant differences from the way corporate shares work. Two important differences that are not immediately apparent are:
a)  SRM guarantees a *minimum* percentage of CPU, rather than a fixed percentage.
b)  Dynamically changed SRM allocations are *not* always immediately reflected in CPU fractions.

Here, the minimum percentage is determined from the ratio of a user's shares to the total shares allocated; something we shall refer to as a user's *entitlement*.

**EXAMPLE 2–1:** Suppose the total CPU allocation on a server is 100 shares. If a particular user has been awarded 10 of those CPU shares by the system administrator, then that user can expect to receive no less than 10% of the CPU resources. A CPU-bound process would consume that percentage. Sometimes, however, it may transpire that the CPU percentage received by that user is higher than the 10% minimum because fewer than the total allocated shares are active. If only 50 of the 100 possible shares are active (because other users and groups are off-line), the active user will see 20% of the CPU. (This is not the way shares work in a corporation.)

Regarding point (b) above, the CPU percentage to which a user is entitled (whether the guaranteed minimum or more), is relative in that it depends on aggregate user activity on the server at any time. As user activity changes, entitlement ratios change, and each user's CPU percentages is modified accordingly. The new entitlements, however, are not immediately reflected in new CPU percentages.

This is because the "instantaneous" CPU usage is only known to the SRM scheduler by sampling per-process usage at a suitably high rate. All advancement or retardation in the frequency with which a process visits the CPU is handled in SRM by adjusting process priorities within the CPU run-queue. In some sense a process is placed incrementally further back or further ahead in the run-queue relative to other processes. Forward migration takes time, so new CPU usage ratios cannot instantly match new entitlements. More details on this in subsequent sections.

**EXAMPLE 2–2:** Suppose 100 total CPU shares are allocated equally to two CPU-bound users on a server. Each SRM user is therefore guaranteed a 50% minimum of the CPU resources. But according to example (2–1) above, the guaranteed minimum only pertains when *both* users are active. If initially, just one user is active then that user will actually receive 100% of the CPU.

When the other user comes online, the 50% SRM minimum now applies but is not immediately realized. Exactly how the guaranteed bound is attained will be discussed shortly. Suffice to say that after some period of time (possibly of order of seconds to minutes) each SRM user will ultimately see the 50% share of the CPU to which they are entitled.

In examples (2–1) and (2–2), we assumed that each user process was CPU-bound for three reasons:
- It represents a form of greed that SRM must throttle under certain conditions.
- Mean CPU utilizations asymptotically reflect SRM resource allocation ratios.
- For capacity planning estimates, it represents the **least upper bound** on resource consumption. (see Figure 1 at end of paper).

We shall adhere to these assumptions for the remainder of this paper, unless otherwise stated.

## 3. SHARES, ENTITLEMENTS and GOALS

We can now formalize the statements from the previous section using the following notation:
- $\$_n$ denotes the **shares** awarded to user n.
- $E_n$ denotes the **entitlement** of user n.
- $U_n$ is the partial CPU **utilization** of user n.

By entitlement we mean the ratio of a particular user's shares to the total shares allocated for CPU resources. On a server with two users having respective allocations $\$_1 = 1$ and $\$_2 = 2$, the entitlement for user 1 is given by:

$$E_1 = \frac{\$_1}{\$_1 + \$_2} \qquad (1)$$

or 0.33 and similarly for user 2, $E_2 = 0.67$.

Next, consider the CPU usage as measured by the SRM scheduler during an arbitrary period T. This is identical to the CPU busy time (B) of each user process [Gunther 98]. In Example (b), with only one user active ($E_1 = 1.0$), the sampled busy time $B_1$ is the same as T for a CPU-bound process i.e., $B_1 = T$. Therefore, the ratio $B_1/T = 1$ which is equivalent to the 100% CPU utilization ($U_1$) seen by user-1. When the second user becomes active, the ratio of entitlements becomes $E_1/E_2 = 0.50$ and the SRM scheduler will adjust each user's CPU usage so that eventually the ratio of partial utilizations reflects that ratio of entitlements:

$$\lim_{T \to \infty} \frac{U_1}{U_2} \sim \frac{E_1}{E_2} \qquad (2)$$

Eqn.(2) is a formal statement of the **goal** (or constraint) that the SRM scheduler aims to achieve. Since entitlement ratios can change dynamically (due to changing user activity), SRM adjusts the partial CPU utilizations to match them. The mathematical limit on the left-hand side means the utilization ratios can only be achieved *in the long run*, rather than instantaneously. Exactly how SRM accomplishes this matching is the subject of the next section.

## 4. CAPACITY PLANNING SCENARIOS

In this section we turn to capacity planning considerations under the operation of the SRM scheduler discussed in [McDougall 1999a and 1999b]. As noted in the Introduction, a system administrator needs guidelines regarding the possible performance impacts when deciding how to allocate SRM shares on a server.

### 4.1 Why Use a Capacity Model?

Because of the dynamics involved, we have chosen to incorporate our understanding of SRM into a performance model using the PDQ© tool provided in [Gunther 1998]. The PDQ Capacity Reports generated by our model can be found at the end of this paper and will be referred to in the subsequent scenarios.

In the following capacity scenarios (1-5), we consider a server that supports three groups of users: a financial database group (FIN), a web server (WEB), and an OPS group (OPS) responsible for support. We treat the FIN and WEB users in aggregate and look at their impact on three explicit users (A, B, C) within the OPS group. We further assume that the CPU service demands (Dcpu) for all users are all equal = 1.

To keep our scenarios suitable for straightforward comparisons, all processes are considered to be CPU-bound (worst case batch demand i.e., thinktime = 0) and only a single process belongs to each user. The reports generated by the PDQ performance model are presented below and are based on the following distribution of 100 shares on the server. Share allocations, entitlements, and the active user shares are shown at the beginning of each table in the format of Table 1.

| Table 1. | Maximum | Share | Allocations |
|---|---|---|---|
| Groups | Allocated | Users | Active |
| FIN | 60 | fAgg | 60 |
| WEB | 10 | wAgg | 10 |
| OPS | 30 | opsA | 6 |
|  |  | opsB | 5 |
|  |  | opsC | 19 |

These parameter values have been selected to provide insight into allocation effects on performance. The PDQ results are easier to compare visually against expected ratios. This is not a limitation of the PDQ/SRM model but rather a limitation of space in this paper which prohibits us from considering all possible parameter combinations.

Under SRM, CPU fractions are proportional to the allocated shares for groups and users within a group. (cf. PDQ Report (4)). These are the least upper bounds mentioned earlier. Table 1 sets the stage for using our PDQ/SRM model to estimate the performance impact of dynamic reassignment of CPU capacity due to a change in the number of active shares.

The key performance metric that we (and most SRM users) are interested in, is Response Time. Response time is the appropriate performance metric for interactive users and transaction-oriented users that have service level targets.

**4.2 Scenario 1 – Two Small-share Users**

**PDQ Capacity Report (1):** We commence with just two users (A and B) active in the OPS group. Everyone else is off-line. PDQ Report (1) shows that the OPS group gets 100% of the CPU split in the ratio 6/11 for userA and 5/11 for userB.

| Table 2. | 2 Active | Users | in OPS |
|---|---|---|---|
| Groups | Allocated | Users | Active |
| FIN | 60 | fAgg | 0 |
| WEB | 10 | wAgg | 0 |
| OPS | 30 | opsA | 6 |
|  |  | opsB | 5 |
|  |  | opsC | 0 |

These CPU fractions are reflected in the corresponding %Ucpu (see the section labeled *Estimated SRM performance* in PDQ Capacity Report (1) at the end of this paper).

There are many different comparisons that can be made using all this data. For brevity, we only consider the following three here:
- Comparing SRM to TS response times within a scenario
- Comparing SRM response times between two successive scenarios
- Comparing SRM response times between any two scenarios

The following tabular format (Table 3) will assist the reader in making those comparisons.

| Table 3. | RT Comparisons in Report 1 | | |
|---|---|---|---|
| User | Rsrm | Rts | Rsrm/Rts |
| opsA | 1.80 | 2.00 | 0.90 |
| opsB | 2.25 | 2.00 | 1.13 |

The column headings have the following meaning. Rsrm is the response time under the SRM scheduler, Rts is the corresponding response time under the conventional Solaris timeshare scheduler. The last column shows the ratio of these two times. If Rsrm / Rts < 1, then SRM can be expected to give better performance than TS. Otherwise, Rsrm/Rts > 1, and SRM can be expected to give worse performance than TS under the selected share allocations.

Because user opsA has 10% more entitlement than she would under a TS scheduler, Under Capacity Scenario 1, her response time is better (Rsrm/Rts < 1) than it would be under TS by 10%, when running CPU-bound. On the other hand, user opsB (having only one less share) suffers more than a 10% degradation in response time relative to that expected under TS.

**4.3 Scenario 2 – Three Active Users**

**PDQ Capacity Report (2):** Now, we consider the impact on response times of bringing user opsC online within the OPS group.

| Table 4. | 3 Active Users in OPS | | |
|---|---|---|---|
| Groups | Allocated | Users | Active |
| FIN | 60 | fAgg | 0 |
| WEB | 10 | wAgg | 0 |
| OPS | 30 | opsA | 6 |
| | | opsB | 5 |
| | | opsC | 19 |

In addition to the columns in Table 3, we append another column to compare the SRM performance ratio between this scenario (2) and the previous scenario (1) i.e., Rs2/ Rs1.

| Table 5. | RT Comparisons in Report 2 | | | |
|---|---|---|---|---|
| User | Rsrm | Rts | Rsrm/Rts | Rs2/Rs1 |
| opsA | 5.34 | 3.00 | 1.78 | 2.97 |
| opsB | 5.80 | 3.00 | 1.93 | 2.58 |
| opsC | 1.56 | 3.00 | 0.52 | N/A |

Because user opsC is entitled to 63% of the CPU, her response time is not only better than the other two OPS users, but twice as good as it would be under TS (Rsrm/Rts = 0.52). However, opsA's response time is now almost three times worse (Rsrm2/Rsrm1 = 2.97) than it was when opsC was absent, and user opsB's performance is about 2.5 times worse (Rsrm2/Rsrm1 = 2.58) than it was in opsC's absence.

Moreover, when user opsC comes online, she is entitled to 63% of the capacity so she is expecting to see her response time less than it would be under TS and she expects to see it improve very smartly!

**4.4 Scenario 3 – Two Active Groups**

**PDQ Capacity Report (3):** Next, we consider the impact on response times of bringing WEB users online. See PDQ Capacity Report (3).

| Table 6. | OPS and WEB Users Active | | |
|---|---|---|---|
| Groups | Allocated | Users | Active |
| FIN | 60 | fAgg | 0 |
| WEB | 10 | wAgg | 10 |
| OPS | 30 | opsA | 6 |
| | | opsB | 5 |
| | | opsC | 19 |

Table 6 shows the group share allocations and the corresponding active shares.

| Table 7. | RT Comparisons in Report 3 | | |
|---|---|---|---|
| User | Rsrm | Rts | Rsrm/Rts | Rs3/Rs2 |
| opsA | 6.97 | 4.00 | 1.74 | 1.31 |
| opsB | 7.88 | 4.00 | 1.97 | 1.36 |
| opsC | 2.08 | 4.00 | 0.52 | 1.33 |

Collecting the response time results in Table 7, we can draw the following striking conclusions. In the presence of the WEB group, all the OPS users suffer between 31% to 36% performance degradation relative to their SRM response times as seen in PDQ scenario 2.

In particular, user opsC suffers a 33% degradation when WEB users becomes active even though she has the greatest CPU entitlement within the OPS group. On the other hand, opsC is still doing almost twice as well (Rsrm/Rts = 0.52) as she would be doing under a TS scheduler.

**4.5 Scenario 4 – All Groups Active**

**PDQ Capacity Report (4):** Next, we consider the impact on response times of bringing FIN users online in Table 8.

| Table 8. | RT Comparisons in Report 4 | | |
|---|---|---|---|
| User | Rsrm | Rts | Rsrm/Rts | Rs4/Rs3 |
| fAgg | 1.67 | 5.00 | N/A | N/A |
| wAgg | 10.0 | 5.00 | 2.00 | 2.50 |
| opsA | 55.6 | 5.00 | 11.1 | 7.98 |
| opsB | 66.8 | 5.00 | 13.4 | 8.47 |
| opsC | 17.6 | 5.00 | 3.52 | 8.46 |

PDQ Report 4 summarizes the expected performance impact. The corresponding allocation table is identical to Table 1. Because FIN owns the majority of the shares, the performance impact on other users is quite significant. See Table 8.

OPS users now suffer a degradation factor of about 8x in response times relative to FIN being off-line in scenario 3. But from the viewpoint of our original pair of OPS users (opsA and opsB) in scenario 1, the news is much worse. They see about a 30x degradation relative to the original configuration in PDQ Report 1.

Note also in the 4th column (Rs4/Rs3) that opsA and opsB are doing worse by more than 10x relative to TS scheduling.

Here is an important point for sysadms. Allowing a sudden and dramatic swing in response times by an order of magnitude or more is highly undesirable and something that a system administrator needs to avoid when allocating SRM shares.

This is a worst case scenario in that we have been looking at the impact of increasingly heavy-share users on the smaller-share users (especially opsA and opsB). Think of this as Goliath's impact on David. We can turn the scenario around and ask, what is David's impact on Goliath? In other words, what is the impact on the heavy-share users of bringing online a lighter weight (but not insignificant) user?

**4.6 Scenario 5 – David's Impact on Goliath**

**PDQ Capacity Report (5):** Even looking at this from the alternative perspective of a single user (e.g., opsC) impacting a group response time like FIN (60% of the shares), we see in Table 9 that WEB and FIN response times can be impacted by as much as 20% degradation.

| Table 9. | RT Comparisons in Report 5 | | | |
|---|---|---|---|---|
| User | Rsrm | Rts | Rsrm/Rts | Rsm/Rs4 |
| fAgg | 1.35 | 4.00 | 0.34 | 0.81 |
| wAgg | 8.10 | 4.00 | 2.03 | 0.81 |
| opsA | 99.4 | 4.00 | 24.9 | 1.79 |
| opsB | 119.5 | 4.00 | 29.9 | 1.79 |

The sysadm needs to be fully aware of any business service-level agreements that may pertain to groups when allocating shares. Although a more benign effect, the impact of an SRM user on a different SRM group may still be very significant when service targets must be taken into account.

**5. CAPACITY PLANNING GUIDELINES**

In the preceding capacity scenarios, we modeled batch-type workloads, considering cpu.shares only with all relative service demands = 1. In some sense, this represents a <u>worst case</u> scenario for expected SRM performance. Your performance will vary (probably less severely) as a consequence of either of the following **ameliorating** influences of:

- your service demands being less relative to our case studies (i.e., D < 1).
- your workload intensity being less relative to our case studies (i.e., Z > 0).

or the following **exacerbating** influences of:
- more than one process per-user relative to our case studies (i.e., N > 1).
- greater disparity in the type of work per group.

**5.1 Top-down Allocations**

In general, the allocation of shares should be done top-down starting with the most important groups, users or known service level targets. They need to be guaranteed a minimum response time. To achieve such performance requirements the following steps are recommended for allocating shares to SRM groups and users:
- Measure their maximum CPU utilization (Umax) under standard TS Solaris.
- Note that the response time (R) is inversely related to Umax.
- Allocate CPU shares such that their entitlement ratio is at least equal to Umax.
- Use the command limadm set cpu.shares = $ to set those shares.
- If these allocations can't be met on as server, use domains, or split groups across separate servers.

**5.2 Metrics to Monitor**

At the moment, none of the generic Solaris performance tools provide the exact performance data that is needed to easily monitor SRM-based performance. The closest that a sysadm can come at present is to monitor and log per-process CPU consumption with a command like /usr/ucb/ps aux or pea.se from the [SEtoolkit].

Recalling that the goal of SRM is expressed in equation (2) in the earlier discussion, the %CPU and TIME fields in the ps command can be used to give some idea of how well SRM is achieving the goal set for it under a top-down allocation of shares.

**6. COMPARISON WITH IBM's SRM & WLM**

Due to space limitations, we can only give the merest hint of the essential differences between Solaris SRM, the older MVS SRM and the newer MVS WLM.

As a point of reference, Solaris SRM now offers resource management functionality that is close in spirit, if not in detail, to the original IBM SRM of more than a decade ago. The IBM version of SRM has changed since its original inception and the descendant is now incorporated into WLM as "compatibility mode" [Samson 97]. WLM now offers "goal mode" as well. UNIX has not taken the key step yet, of defining workloads and service periods and so cannot offer the equivalent of goal mode.

According to eqn(2) in section 3, we can think of Solaris SRM as "looks" at the cumulative CPU (or resource) **busy time** (B). This time is called "CPU Using" in RMF and "cpu usage" in UNIX parlance. The ratio of B over some measurement interval (T) is the steady-state CPU utilization. This CPU utilization, when compared across al tasks or processes, corresponds to the used fraction of CPU resources.

As processes continue to accrue CPU time, they are repositioned in the Solaris run-queue in such a way that their relative ratios asymptotically converge to their respective entitlement ratios in eqn (2).

On the other hand, WLM in Goal Mode "looks" queue-ward at the *waiting time* in the ready-queue and adjusts that time relative to the busy time by repositioning each task in the queue to meet response time or execution velocity objectives in each service class period. This repositioning also takes into account other measures of performance such as *Importance* and the *Performance Index*. For more a more detailed discussion regarding WLM Velocity goals, the interested reader is referred to [Gunther 99a].

In this sense, SRM and WLM "look in opposite directions" to monitor their respective goals. Solaris SRM and MVS WLM act on goals that are compliments of the CPU capacity spectrum. Solaris SRM monitors CPU utilization, whilst MVS WLM-GM monitors delay or wait time.

Obviously, both Solaris SRM and MVS WLM are extremely complex pieces of software and the above analogy is not intended to represent the full functionality. Rather, it is meant to give a system administrator a first-order view of how differently system resource management is controlled in MVS and UNIX and how careful one needs to be when setting SRM or WLM tuning parameters.

## 7. CONCLUSION

Solaris® SRM implements "fair-share" resource allocator through modifications to the traditional UNIX time-share scheduler. In this paper, we have shown how the SRM fair-share scheduler guarantees a least upper bound on CPU consumption for users and groups of users. Because it is a least upper bound, users may actually receive more CPU resource than the bound if fewer than the maximum allocated shares are active. Due to the potential for severe performance swings in response times under such dynamic conditions, it is recommended that fair-share be enabled on servers running workloads where the active allocations are persistent over long periods of time (e.g., batch computations). For online users, service level targets should be carefully scrutinized when allocating shares. Future implementations of SRM can be expected to support service goals.

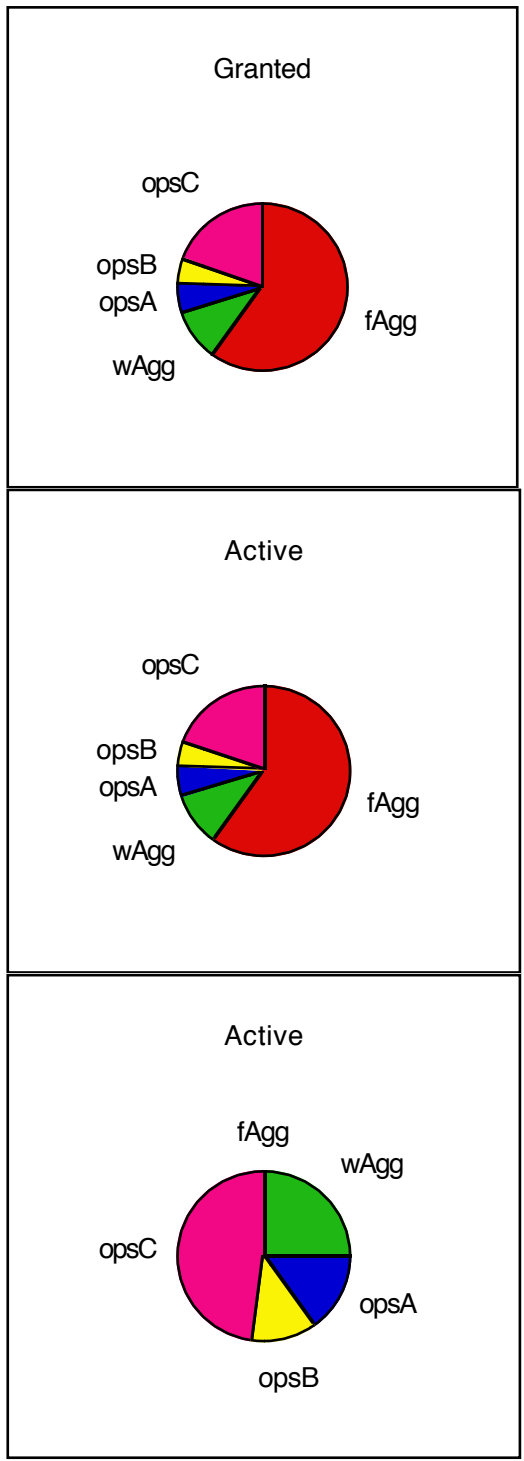

**(A)** This pie chart shows the least upper bound on the CPU fraction allocated across all users under SRM. The groups fAgg and wAgg are treated as aggregate users.

"Least upper bound" means a user is guaranteed to get at least this fraction of CPU resources. Their workload, however, may not be CPU-bound and therefore may never actually consume this smallest guaranteed fraction of CPU.

**(B)** If all users are active and completely CPU-bound (as they are in PDQ Capacity Report (4)), they will consume all of their respective CPU fractions, which correspond identically to their SRM share allocations.

**(C)** If a group of users is inactive (e.g., fAgg in PDQ Capacity Report (3)), then SRM allows the remaining active users to consume a share of the CPU that is proportionately greater than the least upper bound guaranteed in (A). Their concomitant response times will also be better than in case (B).

Figure 1. SRM share allocations compared with different numbers of active shares.

## PDQ Capacity Report (1)

```
Allocations
-----------
11 ACTIVE group cpu.shares out of 100 Allocated.
11 ACTIVE user cpu.shares out of  11 Active group shares.
FIN Group cpu.shares:  60 (offline)
WEB Group cpu.shares:  10 (offline)
OPS Group cpu.shares:  30
       OPS cpu.shares owned by usrA:   6
       OPS cpu.shares owned by usrB:   5
       OPS cpu.shares owned by usrC:  19 (offline)

Group Entitlements
------------------
Group   %Active %UserA %UserB  %UserC
FIN       0.00   0.00          0.00          0.00
WEB       0.00   0.00          0.00          0.00
OPS     100.00  54.55         45.45          0.00

User Workload Parameters
------------------------
User   Procs   Think   Dcpu
opsA    1.00   0.00    1.0000
opsB    1.00   0.00    1.0000

Estimated SRM Performance
-------------------------
User   Thru   RTime   %Ucpu
opsA   0.56   1.80    55.51
opsB   0.44   2.25    44.49

Comparative TS Performance
--------------------------
User   Thru   RTime   %Ucpu
opsA   0.50   2.00    50.00
opsB   0.50   2.00    50.00
```

## PDQ Capacity Report (2)

```
Allocations
-----------
30 ACTIVE group cpu.shares out of 100 Allocated.
30 ACTIVE user cpu.shares out of  30 Active group shares.
FIN Group cpu.shares:  60 (offline)
WEB Group cpu.shares:  10 (offline)
OPS Group cpu.shares:  30
        OPS cpu.shares owned by usrA:    6
        OPS cpu.shares owned by usrB:    5
        OPS cpu.shares owned by usrC:   19

Group Entitlements
------------------
Group  %Active %UserA %UserB %UserC
FIN     0.00         0.00        0.00        0.00
WEB     0.00         0.00        0.00        0.00
OPS   100.00        20.00       16.67       63.33

User Workload Parameters
------------------------
User    Procs   Think   Dcpu
opsA    1.00    0.00    1.0000
opsB    1.00    0.00    1.0000
opsC    1.00    0.00    1.0000

Estimated SRM Performance
-------------------------
User    Thru    RTime   %Ucpu
opsA    0.19    5.34    18.72
opsB    0.17    5.80    17.25
opsC    0.64    1.56    64.02

Comparative TS Performance
--------------------------
User    Thru    RTime   %Ucpu
opsA    0.33    3.00    33.33
opsB    0.33    3.00    33.33
opsC    0.33    3.00    33.33
```

## PDQ Capacity Report (3)

```
Allocations
-----------
40 ACTIVE group cpu.shares out of 100 Allocated.
40 ACTIVE user cpu.shares out of  40 Active group shares.
FIN Group cpu.shares:  60 (offline)
WEB Group cpu.shares:  10
OPS Group cpu.shares:  30
        OPS cpu.shares owned by usrA:   6
        OPS cpu.shares owned by usrB:   5
        OPS cpu.shares owned by usrC:  19

Group Entitlements
------------------
Group  %Active %UserA %UserB  %UserC
FIN     0.00         0.00           0.00           0.00
WEB    25.00        25.00           0.00           0.00
OPS    75.00        15.00          12.50          47.50

User Workload Parameters
------------------------
User    Procs  Think   Dcpu
wAgg     1.00   0.00   1.0000
opsA     1.00   0.00   1.0000
opsB     1.00   0.00   1.0000
opsC     1.00   0.00   1.0000

Estimated SRM Performance
-------------------------
User    Thru    RTime   %Ucpu
wAgg    0.25    4.00    25.00
opsA    0.14    6.97    14.35
opsB    0.13    7.88    12.69
opsC    0.48    2.08    47.96

Comparative TS Performance
--------------------------
User    Thru    RTime   %Ucpu
wAgg    0.25    4.00    25.00
opsA    0.25    4.00    25.00
opsB    0.25    4.00    25.00
opsC    0.25    4.00    25.00
```

**PDQ Capacity Report (4)**

```
Allocations
-----------
100 ACTIVE group cpu.shares out of 100 Allocated.
100 ACTIVE user cpu.shares out of 100 Active group shares.
FIN Group cpu.shares:  60
WEB Group cpu.shares:  10
OPS Group cpu.shares:  30
        OPS cpu.shares owned by usrA:    6
        OPS cpu.shares owned by usrB:    5
        OPS cpu.shares owned by usrC:   19

Group Entitlements
------------------
Group   %Active %UserA  %UserB  %UserC
FIN      60.00   60.00    0.00    0.00
WEB      10.00   10.00    0.00    0.00
OPS      30.00    6.00    5.00   19.00

User Workload Parameters
------------------------
User    Procs   Think   Dcpu
fAgg     1.00    0.00   1.0000
wAgg     1.00    0.00   1.0000
opsA     1.00    0.00   1.0000
opsB     1.00    0.00   1.0000
opsC     1.00    0.00   1.0000

Estimated SRM Performance
-------------------------
User    Thru    RTime   %Ucpu
fAgg     0.60    1.67   60.00
wAgg     0.10   10.00   10.00
opsA     0.02   55.62    5.99
opsB     0.01   66.74    4.99
opsC     0.06   17.60   18.94

Comparative TS Performance
--------------------------
User    Thru    RTime   %Ucpu
Fin      0.20    5.00   20.00
Web      0.20    5.00   20.00
opsA     0.20    5.00   20.00
opsB     0.20    5.00   20.00
opsC     0.20    5.00   20.00
```

**PDQ Capacity Report (5)**

```
Allocations
-----------
81 ACTIVE group cpu.shares out of 100 Allocated.
81 ACTIVE user cpu.shares out of  81 Active group shares.
FIN Group cpu.shares:  60
WEB Group cpu.shares:  10
OPS Group cpu.shares:  30
        OPS cpu.shares owned by usrA:   6
        OPS cpu.shares owned by usrB:   5
        OPS cpu.shares owned by usrC:  19 (offline)

Group Entitlements
------------------
Group   %Active %UserA  %UserB  %UserC
FIN     74.07   74.07    0.00    0.00
WEB     12.35   12.35    0.00    0.00
OPS     13.58    7.41    6.17    0.00

User Workload Parameters
------------------------
User    Procs   Think   Dcpu
fAgg    1.00    0.00    1.0000
wAgg    1.00    0.00    1.0000
opsA    1.00    0.00    1.0000
opsB    1.00    0.00    1.0000

Estimated SRM Performance
-------------------------
User    Thru    RTime   %Ucpu
fAgg    0.74     1.35   74.07
wAgg    0.12     8.10   12.35
opsA    0.01    99.42    7.41
opsB    0.01   119.52    6.16

Comparative TS Performance
--------------------------
User    Thru    RTime   %Ucpu
Fin     0.25     4.00   25.00
Web     0.25     4.00   25.00
opsA    0.25     4.00   25.00
opsB    0.25     4.00   25.00
```